\documentclass[final]{aa}

\usepackage{color}
\usepackage{graphicx}
\usepackage{natbib}
\usepackage{txfonts}

\begin{document}

\title{Probing the variability of the fine-structure constant\\
  with the VLT/UVES%
  \thanks{Based on observations made with ESO Telescopes at the La
    Silla or Paranal Observatories under programme ID 066.A-0212.}}
\author{%
  Ralf Quast\inst{1}
  \and
  Dieter Reimers\inst{1}
  \and
  Sergei A. Levshakov\inst{2}
}
\institute{%
  Hamburger Sternwarte,
  Universit\"at Hamburg,
  Gojenbergsweg 112,
  D-21029 Hamburg,
  Germany
  \and
  Department of Theoretical Astrophysics,
  Ioffe Physico-Technical Institute,
  194021 St.~Petersburg,
  Russia
}
\offprints{R.~Quast,\\ \email{rquast@hs.uni-hamburg.de}}

\date{Received 11 November 2003 / Accepted }


\defcitealias{MurphyFWDP_2003}{MFW}
\defcitealias{MurphyWF_2003}{MWF}

\abstract{%
We assess the cosmological variability of the fine-structure constant
$\alpha$ from the analysis of an ensemble of \ion{Fe}{ii}~$\lambda1608$,
$\lambda2344$, $\lambda2374$, $\lambda2383$, $\lambda2587$, and
$\lambda2600$ absorption lines at the redshift $z=1.15$ toward the
QSO HE~0515--4414 by means of the standard many-multiplet (MM) technique
and its revision based on linear regression (RMM). This is the first
time the MM technique is applied to exceptional high-resolution and high
signal-to-noise QSO spectra recorded with the UV-Visual Echelle
Spectrograph (UVES) at the ESO Very Large Telescope (VLT). Our analysis
results in $\langle\Delta\alpha/\alpha\rangle_\mathrm{MM}=(0.1\pm1.7)
\cdot10^{-6}$ and $\langle\Delta\alpha/\alpha\rangle_\mathrm{RMM}=(-0.4
\pm1.9\pm2.7_\mathrm{sys})\cdot10^{-6}$, which are the most stringent
bounds hitherto infered from an individual QSO absorption system. Our
results support the null hypothesis $\langle\Delta\alpha/\alpha\rangle=0$
at a significance level of 91~percent, whereas the support for the result
$\langle\Delta\alpha/\alpha\rangle=-5.7\cdot10^{-6}$ presented in former
MM studies is 12~percent.
}

\authorrunning{R.~Quast et al.}
\maketitle

\section{Introduction} 
\label{sc:intro}
\enlargethispage{2pt}

Modern 10~m class telescopes equipped with instruments like the
High-Resolution Echelle Spectrograph (HIRES) at the Keck Observatory or
the UV-Visual Echelle Spectrograph (UVES) at the ESO Very Large
Telescope (VLT) facilitate the accurate observation of QSO absorption (or
emission) lines in order to study the hypothetical variability of
fundamental physical constants like the fine-structure constant
$\alpha=e^2/(4\pi\epsilon_0\hbar c)$ or the proton-electron mass ratio
$\mu=m_\mathrm{p}/m_\mathrm{e}$. The interest in these studies is
motivated by the unification theories incorporating varying fundamental
constants \citep[for a review see, e.g.,][]{UzanJ_2003}.

From the astronomical point of view the cosmological variability of the
fine-structure constant is assessed as
\begin{equation}
  \Delta\alpha/\alpha =
    (\alpha_z-\alpha_0)/\alpha_0,
  \label{eq:dalpha}
\end{equation}
where $\alpha_0$ and $\alpha_z$ denote the values of the fine-structure
constant in the laboratory and the specific absorption (or emission)
line system at redshift $z$, respectively. While observational studies
based on the fine-structure splitting of intergalactic alkali-doublet (AD)
absorption lines \citep[e.g.,][]{LevshakovS_1994,MurphyWFPW_2001} and
intrinsic QSO emission line doublets \citep{BahcallSS_2003} have provided
robust upper bounds on $|\Delta\alpha/\alpha|$, \citet[hereafter
MFW]{MurphyFWDP_2003} have recently detected a non-zero expectation value
$\langle\Delta\alpha/\alpha\rangle$ for a sample of 143 complex metal
absorption systems identified in QSO spectra recorded with the Keck/HIRES:
$\langle\Delta\alpha/\alpha\rangle=(-5.7\pm1.1)
\cdot10^{-6}$ in the redshift range $0.2<z<4.2$. This remarkable
statistical evidence for a cosmological variation of the fine-structure
constant is achieved by means of the many-multiplet (MM) technique, which
is a generalization of the AD method incorporating the multi-component
profile decomposition of many transitions from different multiplets of
different ionic species \citep[hereafter MWF, and references
therein]{DzubaFW_1999,MurphyWF_2003}.
While the MM technique considerably improves the formal accuracy,
it also shows some immanent deficiencies. \citet{LevshakovS_2003}
illustrates in detail that the prerequisite assumption that the spatial
distribution is the same for all ionic species is not valid in typical
intergalactic absorption systems. Consequently, it is more reliable to
apply the MM technique to samples of absorption lines arising from only
one ionic species. Furthermore, in comparison to the AD calculations the
incorporation of more transitions over a wider wavelength range results in
a stronger susceptibility to systematic effects. Nevertheless, sources of
error like wavelength miscalibration, spectrograph temperature variations,
atmospheric dispersion, and isotopic or hyperfine-structure effects do
demonstrably not explain the detected non-zero expectation value
\citep[MWF, MFW]{MurphyWFCP_2001}. The observational discrepancy
grows since the radioactive decay rates of certain long-lived nuclei
deduced from geophysical and meteoritic data provide a stringent bound,
$\langle\Delta\alpha/\alpha\rangle=(8\pm8)\cdot10^{-7}$, back to the epoch
of Solar system formation, $z\le0.45$ \citep{OlivePQMV_2003}.
\enlargethispage{2pt}

In the optical spectroscopy, the most striking deficiency inherent to all
decomposition techniques are non-linear inter-parameter correlations
preventing the accurate optimization of the model parameters and possibly
causing ambigous results. In fact, the spectral resolution attained in QSO
observations at the 10~m class telescopes is still not sufficient to
resolve the metal lines with an expected minimum thermal width of about
1~km\,s$^{-1}$. Clearly, in order to solve the profile decomposition
problem spectral observations with the highest possible resolution and
the highest possible signal-to-noise ratio are desirable.

\begin{table}
  \centering
  \caption{Atomic data of the \ion{Fe}{ii} transitions between
    $\lambda1608$ and $\lambda2600$. The laboratory wavelengths $\lambda$,
    oscillator strengths $f$, and relativistic correction coefficients $q$
    are excerpted from \citetalias[Table~2]{MurphyWF_2003} and
    \citet[Table~1]{DzubaFKM_2002}. The sensitivity coefficient $Q$ is
    defined in Sect.~\ref{sc:rmm}. Estimiated errors are indicated in
    parentheses}
  \begin{tabular}{@{}lllll@{}}
    \hline
    \hline
    Tr.\rule[-5pt]{0pt}{15pt}
      & $\lambda$ (\AA)
      & $f$
      & $\phantom{-}q$ (cm$^{-1}$)
      & $\phantom{-}Q$\\
    \hline
    $1608$\rule{0pt}{10pt}
      & $1608.45080\,(8)$
      & $0.058$
      & $-1300\,(300)$
      & $-0.021\,(5)$\\
    $2344$
      & $2344.2130\,(1)$
      & $0.114$
      & $\phantom{-}1210\,(150)$
      & $\phantom{-}0.028\,(4)$\\
    $2374$
      & $2374.4603\,(1)$
      & $0.0313$
      & $\phantom{-}1590\,(150)$
      & $\phantom{-}0.038\,(4)$\\
    $2383$
      & $2382.7642\,(1)$
      & $0.320$
      & $\phantom{-}1460\,(150)$
      & $\phantom{-}0.035\,(4)$\\
    $2587$
      & $2586.6496\,(1)$
      & $0.06918$
      & $\phantom{-}1490\,(150)$
      & $\phantom{-}0.039\,(4)$\\
    $2600$
      & $2600.1725\,(1)$
      & $0.23878$
      & $\phantom{-}1330\,(150)$
      & $\phantom{-}0.035\,(4)$\\
    \hline
  \end{tabular}
  \label{tb:adata}
\end{table}

In this study, we present exceptional high-resolution and high
signal-to-noise spectra of the notably bright intermediate redshift QSO
HE~0515--4414 \citep[$z=1.73$, $B=15.0$]{ReimersHRW_1998} recorded with the
VLT/UVES. The spectra reveal a multi-component complex of metal absorption
lines associated with a sub-damped Lyman-$\alpha$ (sub-DLA) system at the
redshift $z=1.15$ \citep{VargaRTBB_2000,QuastBR_2002,ReimersBQL_2003}. We
analyze a homogenous subsample of \ion{Fe}{ii} $\lambda1608$,
$\lambda2344$, $\lambda2374$, $\lambda2383$, $\lambda2587$, and
$\lambda2600$ absorption lines by means of the standard and a revised MM
technique in order to assess the cosmological variability of the
fine-structure constant.

\section{Observations} 

\begin{figure}
  \centering
  \includegraphics*[94pt,10pt][343pt,310pt]{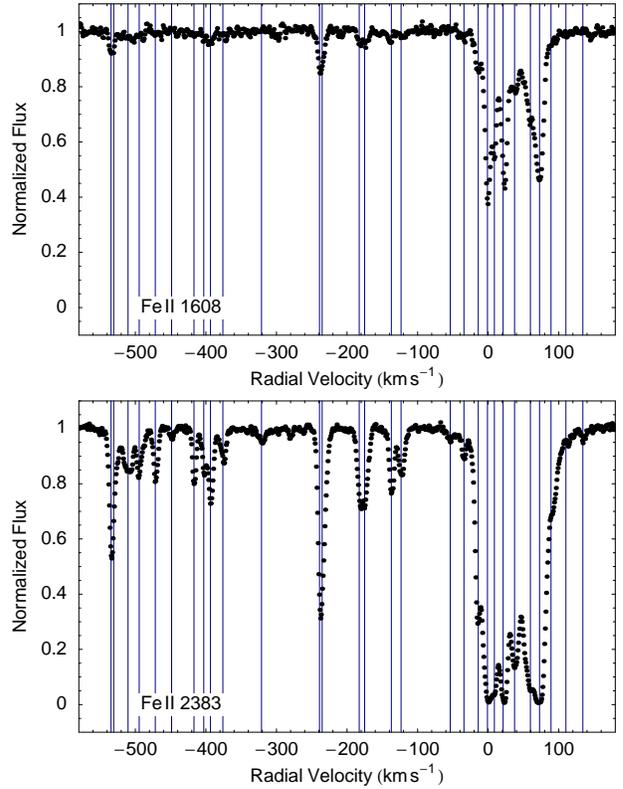}%
  \caption[Total]{Multi-component \ion{Fe}{ii} absorption complex at
    redshift $z=1.15$. For convenience, only the transitions
    $\lambda1608$ and $\lambda2383$ are shown. Individual components
    are marked by a vertical line. The zero point of the radial velocity
    corresponds to the redhsift $z=1.1508$. A close-up ranging from $-20$
    to $100$~km\,s$^{-1}$ is provided in Fig.~\ref{fg:close-up}}
  \label{fg:total}
\end{figure}

HE~0515--4414 was observed with UVES during ten nights between October~7,
2000 and January~3, 2001. Thirteen exposures were made in the dichroic
mode using standard settings for the central wavelenghts of 3460/4370~\AA\
in the blue, and 5800/8600~\AA\ in the red. The CCDs were read out in fast
mode without binning. Individual exposure times were 3600 and 4500~s,
under photometric to clear sky and seeing conditions ranging from 0.47 to
0.70 arcsec. The slit width was 0.8 arcsec providing a spectral resolution
of about 55\,000 in the blue and slightly less in the red. The ThAr lamp
exposures taken immediately after each science exposure provide an
accurate calibration in wavelength. The standard deviation of the
wavelength versus pixel dispersion solution is about 2.0~m\AA\ (2.5~m\AA)
in the blue (red), resulting in an absolute accuracy of about
0.15~km\,s$^{-1}$ in radial velocity space.
The raw data frames were reduced at the ESO Quality Control Garching using
the UVES pipeline Data Reduction Software. The calibrated spectra were
converted to vacuum wavelengths according to \citet{EdlenB_1966} while
the barycentric velocity correction was manually cross-checked using the
ESO-Munich Image Data Analysis Software and the Image Reduction and Analysis
Facility. 
The individual vacuum-barycentric corrected spectra were manually cleaned
from cosmic rays or pixel defects, rescaled to a common median flux level,
and resampled to an equidistant wavelength grid using natural cubic spline
interpolation. The combined spectra (Fig.~\ref{fg:total}) show an effective
signal-to-noise ratio per pixel typically better than 100.

\begin{figure*}
  \centering
  \includegraphics*[94pt,14pt][591pt,320pt]{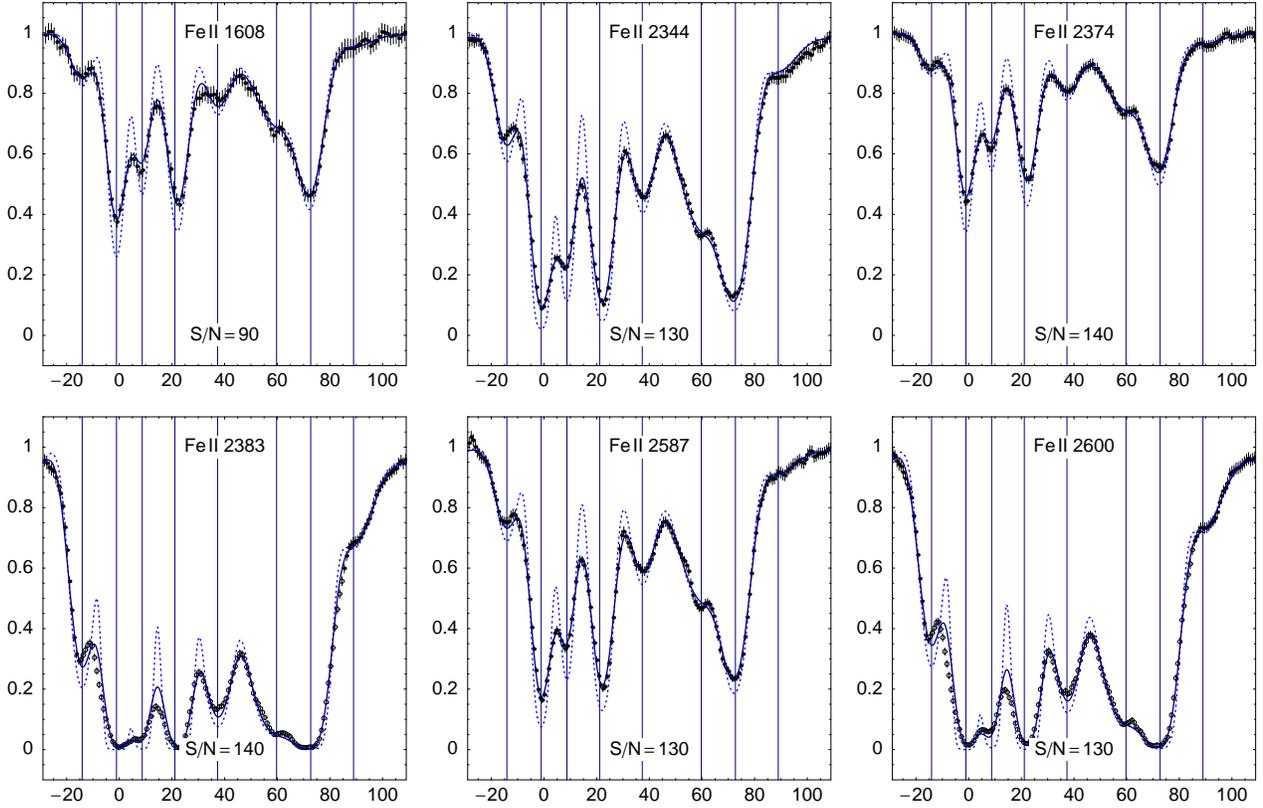}%
  \caption[Close-up]{Close-up of Fig.~\ref{fg:total} showing the subset of
    line profiles associated with the central region of the sub-DLA
    system. The solid and dashed curves represent the optimized model and
    its deconvolution, respectively. The standard deviation of the
    normalized flux is indicated by vertical bars whereas the effective
    signal-to-noise ratio per pixel is stated explicitly. Data marked by
    empty circles are ignored in the optimization}
  \label{fg:close-up}
\end{figure*}

\section{Analysis} 

\subsection{The standard MM technique}
\label{sc:smm}

The observed spectral flux $F(\lambda)$ is modelled as the product of the
background continuum $C(\lambda)$ and the absorption
term convoluted with the instrumental profile, i.e.,
$
  F(\lambda) =
    C(\lambda)\int P(\xi)
      \,\mathrm{e}^{-\tau(\lambda-\xi)}\,\mathrm{d}\xi.
  \label{eq:convolution}
$
While the background continuum is locally approximated by an optimized
linear combination of Legendre polynomials of up to second order, the
instrumental profile is modelled by a normalized Gaussian given by the
the spectral resolution of the instrument. Assuming pure Doppler
broadening, the optical depth $\tau(\lambda)$ is a superposition of
Gaussian functions
\begin{equation}
  g_i(\lambda) =
    \frac{e^2}{4\epsilon_0 mc}
    \frac{N_i f_i \lambda_i}{\sqrt{\pi}b_i}\,
      \exp{
        \left[-
          \left(c\,
            \frac{\lambda-Z_i\lambda_i}{Z_i\lambda_i b_i}
          \right)^2
        \right]},
  \label{eq:gaussian}
\end{equation}
where $Z_i=(1+z_i)(1+u_i/c)$ while $\lambda_i$, $f_i$, $z_i$, $u_i$,
$b_i$, and $N_i$ denote, respectively, the systemic rest wavelength, the
oscillator strength, the cosmological redshift, the radial velocity, the
line broadening velocity, and the column density corresponding to the
line~$i$. The systemic rest wavenumber $\omega_i=1/\lambda_i$ is
parametrized as
\begin{equation}
  \omega_i =
    \omega_{0,i} + q_i\,(\Delta\alpha/\alpha)\,
      (
        2 + \Delta\alpha/\alpha
      ),
  \label{eq:mm}
\end{equation}
where $\omega_{0,i}=1/\lambda_{0,i}$ is the wavenumber in the laboratory
and $q_i$ is the relativistic correction coefficient
\citep{DzubaFKM_2002}.

Even though the sub-DLA system exhibits many additional metal absorption
lines (\ion{Al}{ii}, \ion{Al}{iii}, \ion{Mg}{i}, \ion{Mg}{ii},
\ion{Si}{ii}, \ion{Cr}{ii}, \ion{Ni}{ii}, and \ion{Zn}{ii}) typically
incorporated in the standard MM analysis, we decide to apply this
technique to the \ion{Fe}{ii} transitions only. The restriction to one
ionic species avoids systematic effects if the spatial distribution is not
the same for all species, and reduces systematic effects arising from
isotopic line shifts. In addition, the set of \ion{Fe}{ii} transitions
between $\lambda1608$ and $\lambda2600$ already provides a very sensitive
combination for probing the variability of the fine-structure constant.

Each absorption component is modelled by a superposition of Doppler
profiles with identical radial velocities, widths, and column densities.
In addition, the value of $\Delta\alpha/\alpha$ is confined to be the same
for all components. In order to find the optimal set of parameter values
for both $\tau(\lambda)$ as well as $C(\lambda)$,
the weighted sum of residual squares is minimized by means of an evolution
strategy (ES) based on the concept of covariance matrix adaption
\citep{HansenMK_2003}. We ignore the ensemble of \ion{Fe}{ii}
$\lambda2383$ and $\lambda2600$ lines associated with the central region
of the sub-DLA system since the profiles are saturated and are otherwise
overemphasized in the optimization. 

\subsection{The regression MM (RMM) technique}
\label{sc:rmm}

The standard MM technique can conveniently be revised to avoid the
deficiencies pointed out in Sect.~\ref{sc:intro}. In fact, this
technique is essentially similar to the method developed by
\citet{VarshalovichD_LevshakovS_1993} in order to infer the cosmological
variability of the proton-electron mass ratio $\mu$ from the analysis of
molecular hydrogen absorption lines. Argueing by analogy
\citep{LevshakovS_2003}, in the regime $|\Delta\alpha/\alpha|\ll1$ we
obtain the linear approximation
\begin{equation}
  z_i =
    z_\alpha + \kappa_\alpha Q_i,
  \label{eq:rmm1}
\end{equation}
where $z_i=\lambda_\mathrm{obs}/\lambda_{0,i}$ and $Q_i=q_i/\omega_{0,i}$
denote, respectively, the observed redshift and the sensitivity
coefficient corresponding to the line $i$, and the slope parameter
$\kappa_\alpha$ is given by
\begin{equation}
  \kappa_\alpha =
    -2(1 + z_\alpha)\,(\Delta\alpha/\alpha).
  \label{eq:rmm2}
\end{equation}
If $\Delta\alpha/\alpha$ is non-zero, $z_i$ and $Q_i$ will be correlated
and we will be able to estimate the slope $\kappa_\alpha$ and the
intercept $z_\alpha$ from the linear regression analysis of the position
of the line centroids in an absorption component. The accuracy of the
regression analysis will be improved, if several absorption line samples
are combined. In this case, the regression procedure can be generalized
appropriately:
\begin{equation}
  \zeta_i =
    (\Delta\alpha/\alpha)\,(Q_i-\bar{Q}),
  \label{eq:rmm3}
\end{equation}
where $\bar{Q}$ denotes the mean sensitivity coefficient of the sample,
and~$\zeta_i=(\bar{z}-z_i)/[2(1+\bar{z})]$ is the normalized redshift
while~$\bar{z}$ denotes the mean redshift of an absorption component (in
radial velocity space $\zeta_i=(\bar{u}-u_i)/[2(c+\bar{u})]$).

In order to determine the central position of several selected
\ion{Fe}{ii} lines (see Sect.~\ref{sc:rmmres}) we basically follow the
same strategy as described in Sect.~\ref{sc:smm}. The only differences
are that we do not incorporate Eq.~(\ref{eq:mm}) and do not confine the
radial velocities to be the same for the lines in the selected
components. We point out explicitly that even though we apply a parametric
profile decomposition technique to determine the position of the line
centroids, the RMM analysis is not tied to any specific modelling
technique. In principle, the position of the line centroids can even be
determined without doing any modelling at all
\citep[cf.][]{LevshakovS_2003}. Throughout the analysis we use the atomic
data listed in Table~\ref{tb:adata}.

\begin{table}
  \centering
  \caption[SMM]{Standard MM analysis: optimized values and formal standard
    deviations of the radial velocity $u$, the line broadening velocity
    $b$, and the column density $N$ of the 29 components constituting
    the \ion{Fe}{ii} absorption complex. The optimized value of
    $\langle\Delta\alpha/\alpha\rangle$ is
    $\langle\Delta\alpha/\alpha\rangle_\mathrm{MM}=(0.1\pm1.7)
    \cdot10^{-6}$. The zero point of the radial velocity corresponds to
    the redshift $z=1.1508$}
  \begin{tabular}{@{}llll@{}}
    \hline
    \hline
    No.\rule[-5pt]{0pt}{15pt}
      & $\phantom{-}u$ (km\,s$^{-1}$)
      & $b$ (km\,s$^{-1}$)
      & $\log N$ (cm$^{-2}$)\\
    \hline
    1\rule{0pt}{10pt}
      & $-534.35\pm0.07$
      & $2.07\pm0.16$
      & $12.29\pm0.02$\\
    2
      & $-530.62\pm0.44$
      & $7.26\pm0.41$
      & $12.06\pm0.03$\\
    3
      & $-510.27\pm0.18$
      & $7.92\pm0.36$
      & $12.14\pm0.02$\\
    4
      & $-494.49\pm0.16$
      & $4.86\pm0.32$
      & $11.91\pm0.02$\\
    5
      & $-471.69\pm0.12$
      & $2.99\pm0.20$
      & $11.91\pm0.02$\\
    6
      & $-448.58\pm0.38$
      & $2.21\pm0.87$
      & $11.15\pm0.06$\\
    7
      & $-416.88\pm0.08$
      & $1.76\pm0.20$
      & $11.91\pm0.02$\\
    8
      & $-403.01\pm0.16$
      & $2.19\pm0.37$
      & $11.79\pm0.02$\\
    9
      & $-393.52\pm0.10$
      & $3.39\pm0.22$
      & $12.12\pm0.01$\\
    10
      & $-375.95\pm0.16$
      & $4.38\pm0.22$
      & $11.77\pm0.02$\\
    11
      & $-321.19\pm0.30$
      & $6.98\pm0.58$
      & $11.52\pm0.04$\\
    12
      & $-239.01\pm0.08$
      & $1.66\pm0.11$
      & $12.55\pm0.02$\\
    13
      & $-235.28\pm0.16$
      & $5.15\pm0.15$
      & $12.51\pm0.02$\\
    14
      & $-182.74\pm0.45$
      & $4.02\pm0.42$
      & $12.07\pm0.07$\\
    15
      & $-175.12\pm0.47$
      & $4.92\pm0.40$
      & $12.18\pm0.05$\\
    16
      & $-137.26\pm0.10$
      & $3.11\pm0.18$
      & $12.05\pm0.01$\\
    17
      & $-123.47\pm0.14$
      & $4.47\pm0.20$
      & $12.00\pm0.02$\\
    18
      & $-53.76\pm0.38$
      & $6.04\pm0.59$
      & $11.43\pm0.05$\\
    19
      & $-34.40\pm0.21$
      & $5.81\pm0.36$
      & $11.77\pm0.03$\\
    20
      & $-14.11\pm0.09$
      & $5.09\pm0.10$
      & $12.85\pm0.01$\\
    21
      & $-1.07\pm0.05$
      & $3.76\pm0.06$
      & $13.56\pm0.01$\\
    22
      & $\phantom{-}8.71\pm0.06$
      & $3.46\pm0.10$
      & $13.28\pm0.01$\\
    23
      & $\phantom{-}22.12\pm0.04$
      & $4.53\pm0.06$
      & $13.54\pm0.01$\\
    24
      & $\phantom{-}37.33\pm0.08$
      & $6.04\pm0.14$
      & $13.13\pm0.01$\\
    25
      & $\phantom{-}59.75\pm0.16$
      & $11.3\pm0.2$
      & $13.48\pm0.01$\\
    26
      & $\phantom{-}72.71\pm0.06$
      & $5.69\pm0.08$
      & $13.50\pm0.01$\\
    27
      & $\phantom{-}88.98\pm0.24$
      & $9.16\pm0.52$
      & $12.51\pm0.02$\\
    28
      & $\phantom{-}110.09\pm1.44$
      & $11.4\pm1.7$
      & $11.70\pm0.08$\\
    29
      & $\phantom{-}134.09\pm0.43$
      & $4.08\pm1.00$
      & $11.19\pm0.07$\\
    \hline
  \end{tabular}
  \label{tb:smm}
\end{table}

\section{Results and discussion} 

\subsection{Standard MM analysis}

The optimized values of the model parameters and the standard deviations
provided by the covariance matrix of the ES are listed in
Table~\ref{tb:smm}. The decomposition of the \ion{Fe}{ii} absorption
complex is quite evident (see Figs.~\ref{fg:total} and \ref{fg:close-up}).
This contrasts with the QSO absorption systems considered in former MM
studies where many components are typically unresolved and many line
profiles are saturated \citepalias[see][Figs.~3 and~4]{MurphyWF_2003}.
Clearly, the components with the most accurately defined line centroids
(No.~1, 12, 20--24, 26) are the most important in the MM analysis.
Expectedly, the adequate profile decomposition is reflected in the
formal accuracy of the result:
$\langle\Delta\alpha/\alpha\rangle_\mathrm{MM}=(0.1\pm1.7)\cdot10^{-6}$
is the most stringent formal bound hitherto infered from an individual QSO
absorption system. The best formal accuracy achieved in former MM analyses
\citepalias[see][Table~3]{MurphyWF_2003} is exceeded by a factor of about
three.

\subsection{Regression MM analysis}
\label{sc:rmmres}

\begin{figure}
  \centering
  \includegraphics*[94pt,3pt][344pt,202pt]{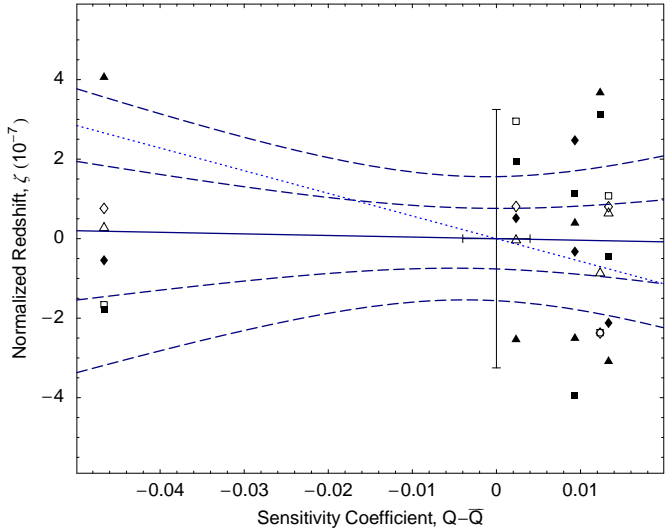}%
  \caption[Regression plot]{Regression MM analysis of the lines listed
    in Table~\ref{tb:rmm}. Each component is represented by a different
    symbol. The regression line
    $\langle\Delta\alpha/\alpha\rangle_\mathrm{RMM}=-0.4\cdot10^{-6}$ and
    its 68 and 95 percent confidence limits are marked by the solid and
    dashed curves, respectively, whereas the dotted line indicates the
    \citetalias{MurphyFWDP_2003} result
    $\langle\Delta\alpha/\alpha\rangle=-5.7\cdot10^{-6}$. The median
    total errors of $\zeta$ and $Q$ are illustrated. }
  \label{fg:rmm}
\end{figure}

The optimized central positions of the lines considered in the RMM
analysis are listed in Table~\ref{tb:rmm}. We do not consider component
No.~22 since the central positions of the lines in this component are
strongly correlated with those in No.~21 and we ignore component
No.~24, because the central position of the \ion{Fe}{ii}~$\lambda1608$
line is not accurately defined.
The regression line (Fig.~\ref{fg:rmm}) indicates no correlation between
the relative displacement of the lines, $\zeta$, and the sensitivity
coefficient $Q$: $\langle\Delta\alpha/\alpha\rangle_\mathrm{RMM}=(-0.4
\pm1.9\pm2.7_\mathrm{sys})\cdot10^{-6}$, where the statistical
errors of $\zeta$ and the estimated errors of $Q$ as well as the
systematic uncertainties inherent to the wavelength calibration are
propagated by means of Monte Carlo simulation. The coefficient of
determination is $r^2=0.005$, i.e. only half a percent of the variation
among the relative displacement of lines is accounted for by the
difference in sensitivity. Consulting the $t$ statistic, the observed
data support the null hypothesis $\langle\Delta\alpha/\alpha\rangle=0$ at
a significance level of 91~percent, whereas the support for the MFW result
$\langle\Delta\alpha/\alpha\rangle=-5.7\cdot10^{-6}$ is 12~percent.

\begin{table*}
  \centering
  \caption[RMM]{Regression MM analysis: optimized centroid positions (\AA)
    of the \ion{Fe}{ii} lines between $\lambda1608$ and $\lambda2600$ in the
    selected absorption components. Formal standard deviations are
    indicated in parentheses}
  \begin{tabular}{@{}lllllll@{}}
    \hline
    \hline
    No.\rule[-5pt]{0pt}{15pt}
      & 1608
      & 2344
      & 2374
      & 2383
      & 2587
      & 2600\\
    \hline
    1\rule{0pt}{10pt}
      & 3453.2908\,(57)
      & 5032.9469\,(24)
      &
      & 5115.7159\,(17)
      & 5553.4531\,(29)
      & 5582.4812\,(19)\\
    12
      & 3456.6993\,(39)
      & 5037.9119\,(23)
      & 5102.9147\,(36)
      & 5120.7677\,(21)
      & 5558.9313\,(28)
      & 5587.9914\,(21)\\
    20
      & 3459.2892\,(37)
      & 5041.6969\,(25)
      & 5106.7434\,(36)
      & 5124.6089\,(22)
      & 5563.1057\,(30)
      & 5592.1855\,(24)\\
    21
      & 3459.4431\,(11)
      & 5041.9145\,(12)
      & 5106.9734\,(12)
      &
      & 5563.3452\,(11)
      &\\
    23
      & 3459.7131\,(13)
      & 5042.3034\,(10)
      & 5107.3695\,(16)
      &
      & 5563.7764\,(11)
      &\\
    26
      & 3460.2938\,(17)
      & 5043.1547\,(15)
      & 5108.2272\,(21)
      &
      & 5564.7129\,(16)
      &\\
    \hline
  \end{tabular}
  \label{tb:rmm}
\end{table*}

\subsection{Systematic effects}

\begin{figure}
  \centering
  \includegraphics{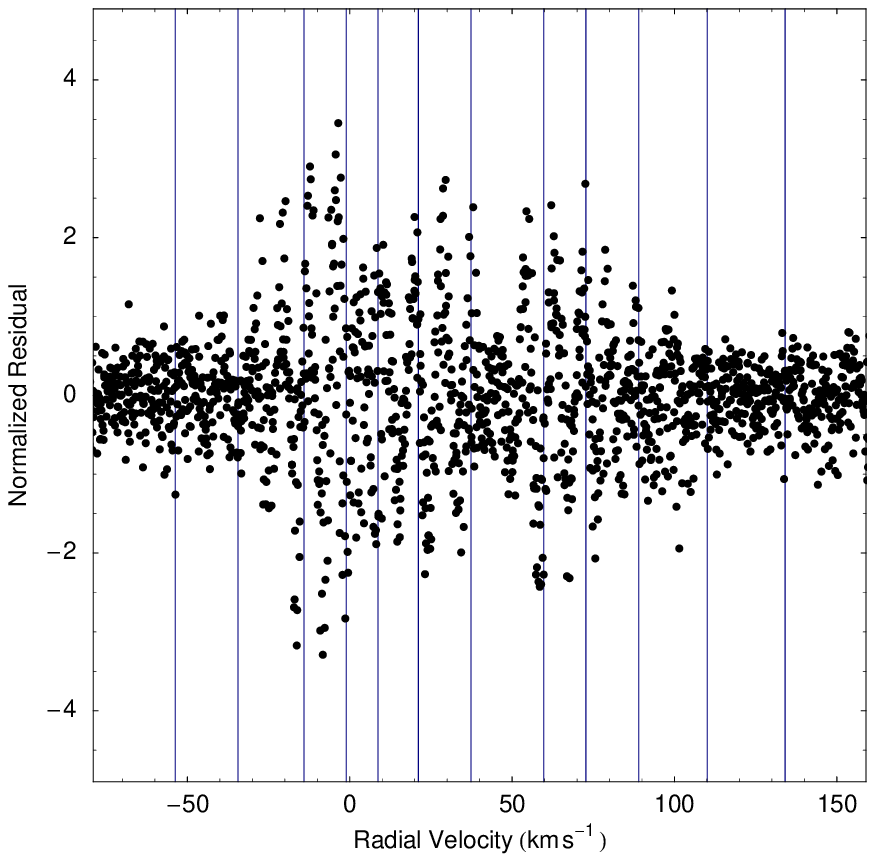}%
  \caption[]{Normalized residuals of the optimized RMM model. For the
    purpose of residual analysis the calculation of the sum of residual
	squares takes into account the optically thick parts of the line
	ensemble. The systematic increase of scatter with increasing optical
	depth may be explained by the presence of unresolved narrow lines
	or by deviations from the Doppler profile resulting from e.g.
	kinematic effects due to rotational motion
	\citep{ProchaskaJ_WolfeA_1997}}
  \label{fg:residuals1}
\end{figure}

\begin{figure}
  \centering
  \includegraphics{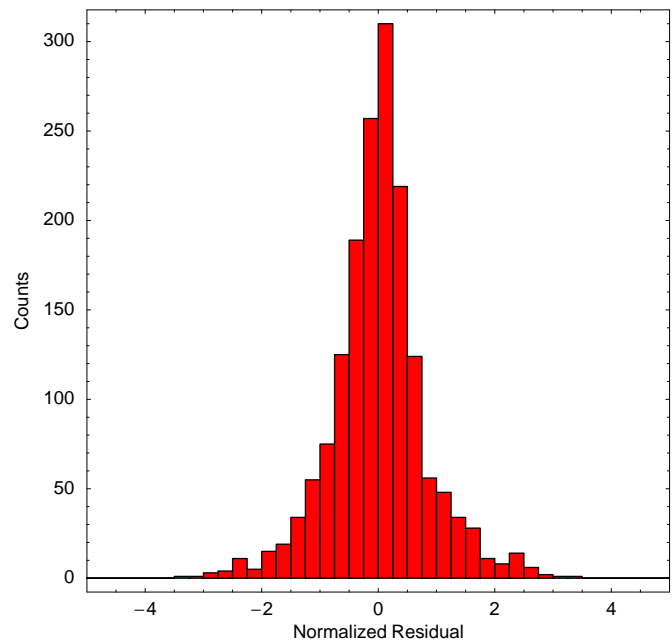}%
  \caption[]{Number distribution of the normalized residuals rendered in
    Fig.~\ref{fg:residuals1}. The distribution is approximately normal
	and does not indicate any missing components}
  \label{fg:residuals2}
\end{figure}

Contributing to the discussion of potential systematic effects presented
in \citetalias{MurphyWF_2003} and \citetalias{MurphyFWDP_2003}, we recall
attention to another important source of error. As formerly illustrated by
\citet{LevshakovS_DodoricoS_1995} and \citet{LevshakovS_1994}, the
presence of unresolved narrow lines with different optical depths can
strongly affect the position of the line centroids in an ensemble of lines
and result in a biased expectation value
$\langle\Delta\alpha/\alpha\rangle$. This effect
will be most noticeable in the case of optically thick lines. In fact, the
expectation value $\langle\Delta\alpha/\alpha\rangle_\mathrm{MM}$
increases by $3.6\cdot10^{-6}$ if we do not ignore the ensemble of
saturated \ion{Fe}{ii} $\lambda2383$ and $\lambda2600$ profiles in the
optimization, and the systematic increase of scatter in the normalized
residuals with increasing optical depth (see Figs.~\ref{fg:residuals1} and
\ref{fg:residuals2} provided in the online material) may be explained by
the presence of unresolved narrow lines.

\section{Conclusions} 

Our results strongly support the null hypothesis of a non-varying
fine-structure constant, but do not contradict the MFW result
$\langle\Delta\alpha/\alpha\rangle=-5.7\cdot10^{-6}$ at a significance
level higher than $88$ percent. Nevertheless, we conclude:
(i)~The MM technique has the capability to provide stringent bounds on
$\langle\Delta\alpha/\alpha\rangle$ even if the analysis is restricted to
the \ion{Fe}{ii} lines only.
(ii)~The RMM technique is illustrative and methodically more transparent
than the standard MM technique. In addition, the regression analysis
facilitates the consideration of systematic errors inherent to the
wavelength calibration of QSO spectra.
(iii)~The accuracy of the individual $\Delta\alpha/\alpha$ assessment is
principally limited by systematic errors inherent to the wavelength
calibration. The accuracy attainable with high-quality QSO spectra
recorded with the VLT/UVES is limited to~$10^{-5}$.
(iv)~Optically thick profiles are susceptible to systematic effects
biasing the expectation value $\langle\Delta\alpha/\alpha\rangle$.
(v)~The analysis of an extensive homogenous sample of \ion{Fe}{ii}
absorption lines is inevitable and will provide an independent and crucial
test of the MFW result.

In particular, HE~0515--4414 is the brightest known intermediate redshift
QSO in the sky and is therefore predestinated for spectroscopy with the
new High Accuracy Radial velocity Planet Searcher (HARPS) operated at
the ESO La Silla 3.6~m telescope. This very high-resolution spectrograph
is specified to provide an efficient wavelength calibration facilitating
the performance of radial velocity measurements with an accuracy of better
than 1~m\,s$^{-1}$ \citep{PepeMR_2002}. Presumably, HARPS will improve
the accuracy of the individual $\Delta\alpha/\alpha$ assessment by more
than an order of magnitude.

\begin{acknowledgements}
  We kindly thank Michael Murphy for valuable comments on the
  manuscript.
  This research has been supported by the Ver\-bund\-forschung of the
  BMBF/DLR under Grant No.~50\,OR\,9911\,1. S.~A.~Levshakov has been
  supported by the RFBR Grant No.~03-02-17522.
\end{acknowledgements}

\bibliographystyle{aa}
\bibliography{aaabbr,ap,rq,sw}

\Online
\onecolumn

\end{document}